\documentclass[apl,twocolumn,showpacs,groupedaddress,preprintnumbers]{revtex4-1}

\usepackage{graphicx}
\usepackage{bm}
\usepackage[mathlines]{lineno}
\usepackage{ulem}
\usepackage{color}
\usepackage{comment}

\begin{document}
\title{Vibrational Energy Transfer from Photo-Excited Carbon Nanotubes to Proteins Observed by Coherent Phonon Spectroscopy}

\author{Tomohito Nakayama$^{1}$\thanks{E-mail: s1720384@s.tsukuba.ac.jp}, Shunsuke Yoshizawa$^{1}$, Atsushi Hirano$^{2}$, 
Takeshi Tanaka$^{2}$, Kentaro Shiraki$^{1}$, and Muneaki Hase$^{1}$\thanks{E-mail: mhase@bk.tsukuba.ac.jp}}

\affiliation{$^{1}$Division of Applied Physics, Faculty of Pure and Applied Sciences, University of Tsukuba, 1-1-1 Tennodai, Tsukuba 305-8573, Japan \\
$^{2}$Nanomaterials Research Institute, National Institute of Advanced Industrial Science and Technology (AIST), Tsukuba Central 5, 
1-1-1 Higashi, Tsukuba 305-8565, Japan}


\begin{abstract}
Vibrational energy transfer from photo-excited single-wall carbon nanotubes (SWCNTs) to coupled proteins is a key to engineer 
thermally induced biological reactions such as photothermal therapy. Here, we explored vibrational energy transfer from the 
photo-excited SWCNTs to different adsorbed biological materials by means of a femtosecond pump-probe technique. We show 
that the vibrational relaxation time of the radial breathing modes (RBMs) in SWCNTs significantly depends on the structure of coupled 
materials, i.e. proteins or biopolymers, indicating the vibrational energy transfer is governed by overlap of phonon density of states 
between the SWCNTs and coupled materials.
\end{abstract}

\maketitle
The energy transfer from mechanical modes to organic layers or specimen has been examined extensively, because of broad interest in efficient energy conversion problems.\cite{Losego12,Hettich16} In particular, one-dimensional and two-dimensional structures made from carbon atoms have shown great advantages on their vivid applications in fundamental systems, such as biological and semiconductor devices,\cite{Modi03,Wang15} because of their extremely high carrier mobility, high thermal conductivity, and rigid structure.\cite{Zhang08} Carbon nanotubes (CNTs) are promising materials for fabricating nano-machines capable for medical treatments,\cite{Kong00} such as photothermal therapy. To realize this potential application, it is highly important to understand protein behaviors on CNTs. Although extensive efforts have been made to understand and control the attraction between CNTs and proteins, only limited knowledge has been obtained on their physical and chemical properties under static conditions.\cite{Horn12}

Single-wall CNTs (SWCNTs) are one of the most popular and fundamental carbon materials, and they have metallic or semiconducting properties,\cite{Saito98} depending on the nanotube symmetry and diameter, which are specified by chiral vector.\cite{Rao97} Frequency-domain spectroscopy revealed diameter-selective Raman scattering from the radial breathing modes (RBMs) in SWCNTs,\cite{Jorio01,Dresselhaus05} where the frequency of the RBMs depends on the inverse of the diameter. Moreover, Hertel {\it et al}. investigated the effects of environment on the optical properties of SWCNTs. They found that interactions between the SWCNTs and their environment depended on their diameter and exciton energy.\cite{Hertel00} Very recently, reduction in antibacterial activity of protein with the help of SWCNTs has been demonstrated.\cite{Horn12} In addition, hen egg-white lysozyme (LYZ) has been demonstrated to disperse SWCNTs, depending on the pH values.\cite{Strano03} However, little is known about the vibrational energy transfer between SWCNTs and proteins, like LYZ, particularly at femtosecond to picosecond time scales, in which phonon-phonon coupling may play a significant role. 
Understanding and controlling the vibrational energy transfer between SWCNTs and proteins (or biopolymers) will be highly relevant to engineering thermally induced biological reactions. 

\begin{figure}
\begin{center}
\includegraphics[width=70mm]{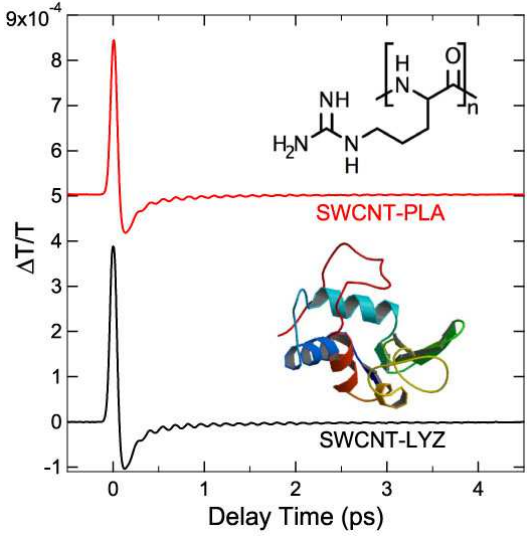}
\end{center}
\caption{Time-resolved transmission obtained for SWCNT-LYZ and SWCNT-PLA at room temperature. The inset shows schematic of the two different absorbed molecules.}
\label{f1}
\end{figure}

In this paper, coherent vibrational motion of the RBMs in SWCNTs coupled with proteins or biopolymers has been precisely measured in the femtosecond time scales. We show that relaxation dynamics of the coherent RBM depend on coupled species, from which it is possible to gain new insights into the vibrational energy transfer from photo-excited SWCNTs to coupled species. 
Here, we selected two kinds of cosolute, i.e. LYZ and poly-L-arginine (PLA), as materials interacting with SWCNTs to investigate their structural effects. Both of them have ability of dispersing SWCNTs, though these solutes have different structures from each other. LYZ consists of 129 amino acid residues (molecular weight: 14,307) with a basic isoelectric point (pI = 11). The tertiary structure of LYZ in aqueous solution is highly stable with secondary structures including $\alpha$-helix and $\beta$-sheet (Fig \ref{f1} inset). On the other hand, PLA (molecule weight: $>$ 70,000) 
is made by polymerizing basic amino acid, arginine (pKa = 12.5) (Fig. \ref{f1} inset). 
As mentioned below, we could observe different dynamics in the coherent RBM phonon between these biomaterials, which is argued to be originated from their structures and thereby phonon density of states (PDOS).

The samples of SWCNTs were produced via the high-pressure catalytic CO decomposition process.\cite{Chiang01} Hen egg-white LYZ and poly-L-arginine (PLA) were obtained from Sigma-Aldrich. The protein solution was prepared as described previously.\cite{Hirano14} Briefly, a solution containing LYZ (0.2 mg/mL) or PLA (0.2 mg/mL) was mixed with the SWCNT powder. Then the SWCNT solution was dispersed by ultrasonication for 60 minutes at 18$^{\circ}$C using ultrasonic processor. 
Finally, highly dispersed SWCNT solution was obtained from supernatant of the sample solution after ultracentrifugation (210,000 g for 30 min at 25$^{\circ}$C) using an ultracentrifuge. 
On the other hand, the bundled sample was also obtained from supernatant of the dispersed solution after gentle centrifugation (16,100 g for 1 minute at 25$^{\circ}$C).

The transient transmission ($\Delta T/T$) of the samples, i.e. SWCNT dispersed by LYZ (SWCNT-LYZ) or by PLA (SWCNT-PLA), was measured in a 5-mm-thick quartz cell by employing the fast scanning delay method.\cite{Hase03} The pump and probe pulses (20 fs pulse duration; 830 nm wavelength; 80 MHz repetition rate) were focused to a 70-$\mu$m-diameter spot in the sample with a mutual angle of 
$\approx$10$^{\circ}$. The pump pulse with a constant average power (40 mW) generated a carrier density of 4 $\times$ 10$^{19}$ cm$^{-3}$; the probe power was less than 2 mW. Under the constant pump power and the focused spot size, the pump fluence should be constant through the measurements, and therefore the effect of excitation density on the vibrational relaxation dynamics, i.e. carrier-density-dependent carrier-phonon scattering, was negligible in the present study. Polarization of the two beams was linear with a mutual angle of 55$^{\circ}$ (magic angle).\cite{Makino09} 
 
The $\Delta T/T$ signals were recorded as a function of the pump-probe time delay as shown in Fig. \ref{f1}. The transient electronic response is observed just at the zero time delay, which is followed by the coherent phonon oscillations due to the RBMs, whose frequency ($\omega_{RBM}$) inversely depends on the diameter of SWCNTs ($R_{CNT}$),\cite{Makino09}
\begin{equation} \label{eq1}
\omega_{RBM} = 7.44 [THz]/R_{CNT} [nm].
\end{equation}

\begin{figure}
\begin{center}
\includegraphics[width=85mm]{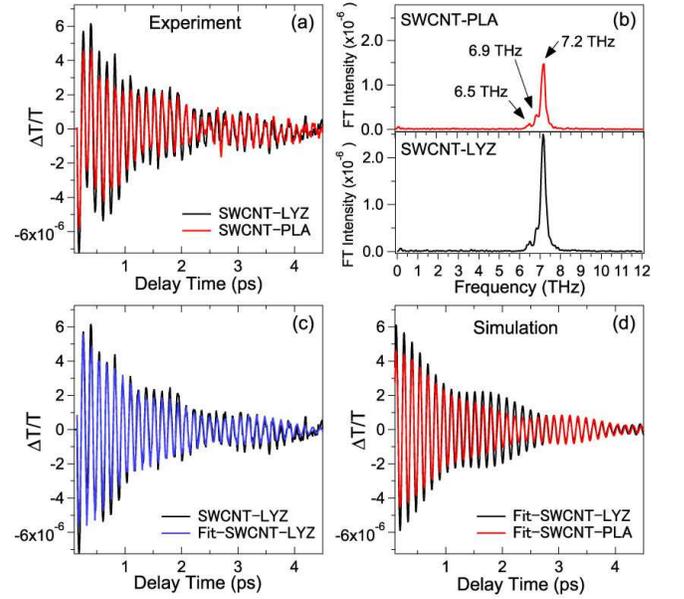}
\end{center}
\caption{(a) Coherent oscillation signals for SWCNT-LYZ (Black) and SWCNT-PLA (Red). (b) Fourier transformed spectra obtained from the experimental time-domain data exhibited in the panel (a). (c) Time-domain data of $\Delta T/T$ (Black) and fitting data (Blue) for SWCNT-LYZ. (d) The simulated data drawn by using eq. (\ref{eq2}) with the parameters listed in Table \ref{t1}.}
\label{f2}
\end{figure}

\begin{figure}
\begin{center}
\includegraphics[width=70mm]{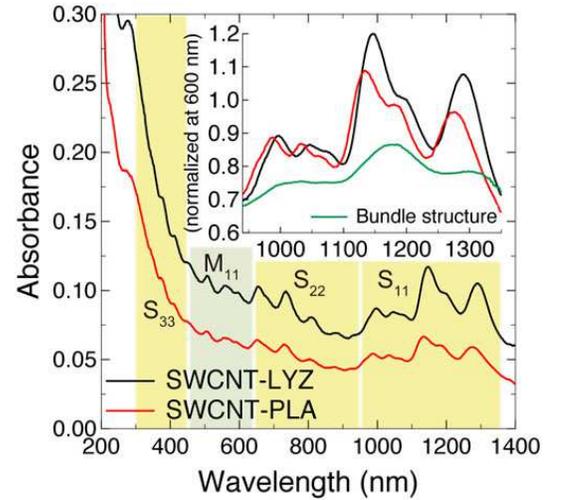}
\end{center}
\caption{Absorption spectra of SWCNT-LYZ (Black) and SWCNT-PLA (Red) dispersions. The inset exhibits the enlarged spectra normalized at 600 nm together with the bundled form of the SWCNT-PLA (Green). 
M$_{11}$ represents the 1st-order metallic optical transition, while S$_{11}$, S$_{22}$, and S$_{33}$ represent the 1st-, the 2nd-  and the 3rd-order semiconducting optical transitions, respectively.\cite{Tanaka09}}
\label{f3}
\end{figure}

The transient electronic response, appearing just at the zero time delay, corresponds to the coherent arrangement of the SWCNTs molecules induced by the electric field of the femtosecond laser pulse, 
and is governed by the third-order nonlinear susceptibility $\chi^{(3)}$ [ref. \cite{Shen84}]. 
To subtract the transient electronic response, we fitted the time domain data (Fig. \ref{f1}) with a double-exponential-decay function. Subsequently, we subtracted the double-exponential-decay function from the time-domain data. Thus, we obtained only oscillatory parts of the time-domain signals (Fig. \ref{f2}). Comparing the $\Delta T/T$ signal obtained in the different systems, i.e., SWCNT-LYZ and SWCNT-PLA, the coherent phonon spectra are indeed different from each other, indicating the interaction between SWCNTs and biomaterials depends on their structures [Fig. \ref{f2}(a)]. By using the Kataura's plot, it is possible to derive 
the electronic states of SWCNTs and hence to identify whether they are semiconducting or metallic. In our samples, the diameter of SWCNTs has a range of 0.9-1.3 nm. 
Considering the photon energy (830 nm, i.e., 1.49 eV) and the broad bandwidth ($\approx$50 nm in FWHM) of the laser used, the excitonic transitions around the critical 
point of S$_{22}$ (the second-order semiconducting optical transition) are available at 800-900 nm [ref. \cite{Kataura99}]. We can thus conclude that the semiconducting 
SWCNTs are resonantly excited through the impulsive stimulated Raman scattering.\cite{Lim06,Gambetta06} 
Actually, the peaks at 6.5, 6.9, and 7.2 THz observed in the Fourier transformed (FT) spectra [Fig. \ref{f2}(b)] are assigned to the 1.14-, 1.08-, and 1.03-nm-diameter tubes, respectively, based on eq. (\ref{eq1}). 
Note that chirality (n, m) of these modes can be (9, 7), (8, 7), and (12, 1) for the 1.14-, 1.08-, and 1.03-nm-diameter tubes, respectively.\cite{Goupalov06}
The time-domain data were fit to a linear combination of damped harmonic oscillators, 
\begin{eqnarray} \label{eq2}
\Delta T/T = A_{1}e^{-t/\tau_{1}}\cos(2 \pi \nu_{1}t + \phi_{1}) 
\nonumber \\
+ A_{2}e^{-t/\tau_{2}}\cos(2 \pi \nu_{2}t + \phi_{2}) \nonumber \\
+ A_{3}e^{-t/\tau_{3}}\cos(2 \pi \nu_{3}t + \phi_{3}),
\end{eqnarray}
where $A$ ($A_{1}$, $A_{2}$, $A_{3}$), $\tau$ ($\tau_{1}$, $\tau_{2}$, $\tau_{3}$), $\nu$ ($\nu_{1}$, $\nu_{2}$, $\nu_{3}$), and $\phi$ ($\phi_{1}$, $\phi_{2}$, $\phi_{3}$) are the amplitude, the relaxation time, the frequency, and the initial phase of the three coherent RBMs of the first ($\nu_{1}$ = 6.5 THz), the second ($\nu_{2}$ = 6.9 THz), and the third ($\nu_{3}$ = 7.2 THz) modes, respectively. 
The time-domain data are well fit to eq. (\ref{eq2}) as shown in Fig. \ref{f2}(c), and the fitting parameters obtained are listed in Table \ref{t1}. By comparing the amplitude of the dominant RBM (7.2 THz) for two biomaterials, we found that the initial vibrational amplitude for the SWCNT-LYZ system is $\approx$1.6 times larger than that for the SWCNT-PLA system (see Table \ref{t1} and Fig. \ref{f2}(d)), which is dominantly ascribable to the difference in the concentration of SWCNTs between these samples\cite{Tanaka09}; note that the absorbance of the SWCNT-LYZ at a 810 nm peak after the background subtraction was $\approx$2.1-fold higher than that of the SWCNT-PLA (Fig. \ref{f3}). 

Here, we focus on the strongest RBM at 7.2 THz by extracting the vibrational energy transfer dynamics from the fitting results. Importantly, the vibrational relaxation time for the SWCNT-LYZ (ca. 1.41 ps) at 7.2 THz was shorter than that for the SWCNT-PLA (ca. 1.66 ps), suggesting that the vibrational energy transfer from the photo-excited SWCNT to LYZ is faster than that from SWCNT to PLA for the 7.2 THz mode. These vibrational relaxation time constants are significantly shorter than those obtained for the SWCNTs dispersed by sodium dodecyl sulfate (SWCNT-SDS) (ca. 2.0 ps), which was observed by the same pump-probe technique,\cite{Makino09} indicating that the vibrational relaxation time of the coherent RBM in the present complex systems was indeed changed by the adsorbed biomaterials. Although decay of the coherent RBM is generally accounted for by anharmonic phonon-phonon energy relaxation into underlying acoustic phonons,\cite{Makino09} this decay path depends only on the temperature, so that the change in the relaxation time observed in Fig. \ref{f2} at the constant room temperature cannot be accounted for by the anharmonic phonon-phonon energy relaxation. 
Note that the total relaxation rate 1/$\tau$ is the sum of the new path and the existing path (anharmonic phonon-phonon energy relaxation), 1/$\tau$ = 1/$\tau_{anharmonic}$+1/$\tau_{new}$, where we will discuss the origin of the new path ($\tau_{new}$) in details below.

It should be also noted that LYZ is adsorbed onto the SWCNTs through hydrophobic and $\pi-\pi$ interactions\cite{Calvaresi12} whereas PLA is adsorbed onto the SWCNTs 
through hydrophobic and van der Waals interactions and partly through $\pi-\pi$ interaction between the guanidium group of arginine and the SWCNT surfaces.\cite{Hirano14} Because of the different adsorption mechanisms, the coverage of the SWCNTs will not be the same for the two types of cosolute. The difference in the coverage of the SWCNTs can be reflected in difference in carrier doping of SWCNTs.\cite{Hirano16} However, in terms of the carrier doping, the differences in the coverage will not affect the vibrational relaxation time of the SWCNTs, since we have already demonstrated that the carrier doping did not affect the relaxation time of the RBMs.\cite{Makino09} 

Since SWCNTs dispersed in aqueous solution generally exist as debundled or bundled forms,\cite{Hirano14} change in the relaxation time might be associated with the intertube interactions. In fact, the bundling effects on the phonon relaxation have been examined using a SWCNT film and SWCNT solution for the $G$-mode\cite{Lee10} as well as for the RBM,\cite{Honda13} indicating that the bundling makes phonon relaxation time shorter than that of individually dispersed SWCNTs. However, our samples used here were individually dispersed since they were collected from the supernatants after the ultracentrifugation. To test this expectation, we measured the absorption spectra of the SWCNTs dispersed solutions as well as the bundled form (Fig. \ref{f3}). 
Although the absorption intensity of the S$_{11}$ band was found to be slightly smaller and broader for the SWCNT-PLA than that for the SWCNT-LYZ, the intensity of the S$_{11}$ band was explicitly stronger than that of the bundled form (see the inset of Fig. \ref{f3}).\cite{Oconnell02} These results indicate that both the SWCNT-LYZ and SWCNT-PLA were well dispersed. Thus, the decay-time shortening of the RBMs seen with LYZ can be explained mainly by the energy transfer from the SWCNTs to LYZ and not by the bundling effect.

\begin{figure}
\begin{center}
\includegraphics[width=80mm]{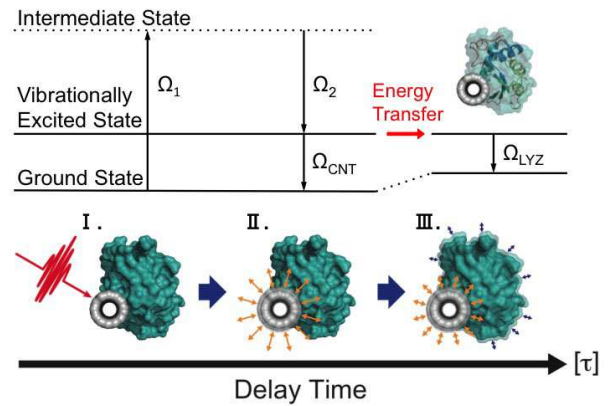}
\end{center}
\caption{Schematic dynamics of the vibrational energy transfer from the SWCNT to LYZ at 7.2 THz. 
The coherent RBM is excited via 
ISRS,\cite{Lim06} whose spectra cover the excited phonon energy ($\Omega_{CNT}$ = $\Omega_{1} - \Omega_{2}$). Because the energy level of $\Omega_{CNT}$ and that of the phonon in LYZ ($\Omega_{LYZ}$) overlap each other, effective energy transfer occurs. 
Three images in (I)--(III) portray the process of phonon-phonon coupling between the SWCNT and LYZ as a function of the delay time. (I), the SWCNT is 
excited. (II), the coherent RBM in the SWCNT starts to oscillate. (III), the vibrational energy of the SWCNT's RBM propagates to the LYZ phonon mode. }
\label{f4}
\end{figure}

To further discuss the difference in the vibrational energy transfer observed in Fig. \ref{f2}, the phonon-phonon coupling between the SWCNTs and biopolymers should be taken into account.\cite{Leitner01,Hase02} 
Recently Li {\it et al}. observed anisotropic energy flow from proteins to ligands on albumin molecules using ultrafast infrared vibrational spectroscopy.\cite{Magana14} The efficient vibrational energy transfer was thought to occur through the connecting helix structures. In SWCNT-protein systems, similarly, favorable coupling would occur between $\alpha$-helix of proteins and SWCNT.\cite{Hu16} 
Therefore, it is expected that the vibrational energy transfer from SWCNTs to coupled LYZ might be more efficient than that from the SWCNT to PLA, since only LYZ has $\alpha$-helix structure. 
We thus examine the vibrational energy transfer from SWCNTs to coupled biomaterials in terms of 
phonon-phonon coupling based on the PDOS.\cite{Leitner01} 
The main peak in the PDOS at $\approx$240 cm$^{-1}$ (= 7.2 THz) for LYZ\cite{Svanidze07} is well overlapped with the energy of the RBM (7.2 THz), while the overlap of the main peak ($\approx$250 cm$^{-1}$ = 7.5 THz) in the PDOS for PLA with the RBM (7.2 THz) is poor.\cite{Sharma02} 
Therefore, there can exist resonant coupling between the RBM (7.2 THz) in the SWCNTs and the optical phonons in LYZ through $\alpha$-helix of proteins, leading to efficient vibrational energy transfer. 
A possible scenario for the vibrational relaxation dynamics in the SWCNT-LYZ complex at 7.2 THz is schematically presented in Fig. \ref{f4}, which exhibits that the vibrational amplitude of the SWCNT readily decays by the vibrational energy transfer to LYZ. 

To further support our interpretation, we then focused on the vibrational relaxation time of the 6.9 THz mode. 
As shown in Table \ref{t1}, the relaxation time for SWCNT-PLA (ca. 2.00 ps) was significantly shorter than that for SWCNT-LYZ (ca. 3.76 ps)
, both of which are shorter than that of the 6.9 THz RBM in SWCNT-SDS (5.50 $\pm$ 0.17 ps)(data not shown). 
This result suggests that the vibrational energy transfer from the photo-excited SWCNT to PLA is faster than that from SWCNT to LYZ, which is concordant with the fact that the peak in the PDOS for PLA ($\approx$230 cm$^{-1}$ = 6.9 THz) well overlaps with the energy of the RBM (6.9 THz).\cite{Sharma02} Importantly, the difference in the relaxation time of the 7.2 THz mode between LYZ and PLA is very small (i.e. 1.41 and 1.66 ps, respectively), while that of the 6.9 THz mode is relatively large (i.e. 3.76 and 2.00 ps, respectively). A plausible reason for this lies in the natural linewidth of the PDOS inside a protein or biopolymer; specifically, the linewidth of the 240 cm$^{-1}$ (7.2 THz) mode for LYZ ($>$ 10 cm$^{-1}$)\cite{Svanidze07} is much broader than that of the 230 cm$^{-1}$ (6.9 THz) mode for PLA ($<$ 10 cm$^{-1}$).\cite{Sharma02}. Therefore, the resonant energy transfer 
will be more pronounced for PLA than for LYZ at the respective peak position in PDOS. This could explain why the difference in the relaxation time is larger in the case of the 6.9 THz mode. According to the above results and discussions, we concluded that the rate of the phonon energy flow significantly depends on the PDOS of the biomaterials. 

\begin{table}
  \caption{The fitting parameters obtained using eq. (\ref{eq2}). The values for $\nu$ were obtained from the peak frequencies in the FT spectra [Fig. \ref{f2}(b)] and treated as the constants during the fitting. The standard deviation of the coefficients were obtained during the fitting procedure with Igor Pro.}
\label{t1}
  \begin{tabular}{cccccc}
  \hline
     Sample & Mode & $\nu$ (THz) & $A \times$10$^{6}$ & $\tau$ (ps) & $\phi$ (deg.)\\
    \hline
     SWCNT  & 1st & 6.5 & 0.59$\pm$ 0.05 & 2.16 $\pm$ 0.20 & -109.1 $\pm$ 2.5 \\
     -LYZ & 2nd & 6.9 & 0.39$\pm$ 0.04 & 3.76 $\pm$ 0.50 & 74.8 $\pm$ 3.2 \\
       & 3rd & 7.2 & 6.38 $\pm$ 0.06 & 1.41 $\pm$ 0.01 & 23.4 $\pm$ 0.3 \\
        \hline
     SWCNT & 1st  & 6.5 & 0.54 $\pm$ 0.06 & 1.94 $\pm$ 0.22 & -130.4 $\pm$ 3.4 \\
     -PLA & 2nd & 6.9 & 0.78 $\pm$ 0.08 & 2.00 $\pm$ 0.17 & 58.6 $\pm$ 2.9 \\
       & 3rd & 7.2 & 4.03 $\pm$ 0.06 & 1.66 $\pm$ 0.03 & 18.5 $\pm$ 0.4 \\
            \hline
  \end{tabular}
\end{table}

In summary, we explored the vibrational energy transfer between the two different biomaterials and SWCNTs by means of the femtosecond transient transmission technique. The vibrational relaxation time of the RBMs dramatically changes, depending on the PDOS of the coupled biomaterials. Our findings are particularly useful for designing a highly efficient phonon energy flow system from photo-excited SWCNT to biomaterials, and such vibrational energy transfer can be controlled by the PDOS originated from the structure of coupled biomaterials. 

The authors thank S. Hasegawa and H. Tadokoro for the assistance at the early stage of the experiments. This work was supported in part by KAKENHI-23104502,  
from MEXT, Japan. 


\end{document}